\documentclass[superscriptaddress,showkeys,twocolumn,nofootinbib,longbibliography]{revtex4-1}

\usepackage{amsmath}
\usepackage{amssymb}
\usepackage{graphicx}
\usepackage{epsf}
\usepackage{slashed}
\usepackage{enumitem}
\usepackage[czech,english]{babel}
\usepackage[cp1250]{inputenc}
\usepackage{hyperref}
\usepackage{xcolor}

\usepackage[only,llbracket,rrbracket,llparenthesis,rrparenthesis]{stmaryrd} % for comparison with 4 similar parenthesis symbols
\usepackage{accsupp} % for ensuring the right Unicode codepoint upon pasting

\newcommand{\be}{\begin{equation}}
\newcommand{\ee}{\end{equation}}

\newcommand{\bea}{\begin{eqnarray}}
\newcommand{\eea}{\end{eqnarray}}

\usepackage{bbm} %\mathbbm{1}, \mathbbmss{1}, \mathbbmtt{1}

\usepackage[nodayofweek]{datetime}
\newdateformat{mydate}{\twodigit{\THEDAY}{ }\shortmonthname[\THEMONTH], \THEYEAR}
\usepackage[normalem]{ulem}
\usepackage{color}
\usepackage{ulem} % podtrhavani, skrtani atd. textu \uline,\uuline,\uwave,\sout,\xout

\definecolor{myred}{rgb}{1,0,0}
\definecolor{mygreen}{rgb}{0,0.8,0.2}
\definecolor{myblue}{rgb}{0,0,1}

\definecolor{Ared}{rgb}{1,0.7,0}
\definecolor{Agreen}{rgb}{0.7,0.8,0.2}
\definecolor{Ablue}{rgb}{0,0.7,1}

\renewcommand{\emph}[1]{\textit{#1}}

\begin{document}

\definecolor{mygreen}{HTML}{006E28}
\newcommand{\kasia}[1]{{\color{mygreen}\textbf{?KS:}  #1}}
\newcommand{\trom}[1]{{\color{red}\textbf{?TR:} \color{red} #1}}
\newcommand{\filip}[1]{\textbf{?FB:} {\color{blue} #1}}
\newcommand{\aw}[1]{{\color{magenta}\textbf{?AW:}  #1}}
%\preprint{APS/123-QED}

 \title{Moduli space metric of the excited vortex}
 
\author{D. Migu\'elez-Caballero}
\affiliation{Departamento de F\'isica Teorica, Ato\'mica y \'Optica and Laboratory for Disruptive Interdisciplinary Science (LaDIS), Universidad de Valladolid, 47011 Valladolid, Spain}

\author{S. Navarro-Obreg\'on}
\affiliation{Departamento de F\'isica Teorica, Ato\'mica y \'Optica and Laboratory for Disruptive Interdisciplinary Science (LaDIS), Universidad de Valladolid, 47011 Valladolid, Spain}

\author{A. Wereszczynski}
%\affiliation{Departamento de Matematica Aplicada,  University of Salamanca, SPAIN}
\affiliation{Institute of Theoretical Physics, Jagiellonian University,
Lojasiewicza 11, Krak\'{o}w, Poland}
\affiliation{International Institute for Sustainability with Knotted Chiral Meta Matter (WPI-SKCM2), Higashi-Hiroshima, Hiroshima 739-8526, JAPAN}

\begin{abstract}
We show that the moduli space metric of a single vortex gets corrections because of the excitation of the radially symmetric shape mode. It leads to a non-zero amount of the shape mode carried by the vortex when moving with a constant velocity. However, due to the radial symmetry, this effect does not reproduce the Lorentz contraction of the moving vortex.  We apply the Derrick mode approximation in order to recover the Lorentz contraction on the level of the collective model.
\end{abstract}

\maketitle

%%%%%%%%%%%%%%%%%%%%%%%%%%%%%%%%%%
\section{Introduction}
%%%%%%%%%%%%%%%%%%%%%%%%%%%%%%%%%%
 Vortices in the Abelian-Higgs model in (2+1) dimensions are prototypical examples of topological solitons with local gauge invariance \cite{NO,Manton2002}. 
They appear as a lower-dimensional (planar) truncation of electroweak theory. If extended along the third dimension, many applications in cosmology and astrophysics can be found; see, e.g., cosmic strings \cite{Vilenkin2000,Manton2002}. 

An especially interesting case is the critical coupling limit at which there is no static vortex-vortex force. In fact, in this regime, known as the BPS limit,  vortices can be found as solutions of the relevant Bogomolny equations, which are static equations of lower order than the Euler-Lagrange equations of motion \cite{BOG}. This allows semi-analytical insight into the static \cite{T}, and more importantly, the dynamical properties of the BPS vortices \cite{S}. In particular, multivortex scattering can be understood as a flow through the available energetically degenerate BPS solutions in the topologically fixed sector \cite{Samols}. These solutions form a space spanned by the so-called moduli, i.e., continuous parameters (collective coordinates) parameterizing the BPS solutions, and equipped with a natural metric \cite{NM}. Each parameter corresponds to a zero mode, that is, a massless excitation of such a multi-vortex solution. The actual dynamics is a force-free motion governed by the moduli space metric. Probably, the most well known result is the $90^\circ$ scattering of two unit charge BPS vortices \cite{Samols, SR, Ruback, Reb}. 

Outside the critical regime, the concept of moduli space is still useful. The dynamics is modified by the appearance of a static force between the vortices, which is attractive (type I) or repulsive (type II), depending on the value of the coupling constant $\lambda$ \cite{Sp}. 

Surprisingly, the dynamics of vortices, even at critical coupling, is much more complex and interesting than that of the standard geodesic flow. The reason is that vortices host massive normal (vibrational) modes \cite{AGM}. Importantly, even though the energy of the BPS solutions does not depend on the value of the moduli (position on the moduli space), the structure of the vibrational modes does \cite{AGMM, AMMRW}. As a consequence, there exists potential energy due to the excitation of the mode, which in the harmonic oscillator approximation depends quadratically on the frequency of the mode \cite{AMMW}. Since the frequency changes on the moduli space, a mode-generated force appears. This has a tremendous impact on vortex scattering \cite{KRW, AMRW,AMMRW}. 

The modes give rise not only to the effective potential but also simultaneously affect the metric of the moduli space. This has a smaller effect on the dynamics than on the appearance of the mode-generated force, but it can still be important for prediction of the actual final state. The aim of the present paper is to understand how the moduli space metric of the single vortex changes by including the normal modes. 
%%%%%%%%%%%%%%%%%%%%%%%%%%%%%%%%%%
\section{Warming up example - the wobbling kink}
%%%%%%%%%%%%%%%%%%%%%%%%%%%%%%%%%%
We begin with a simple example, which is the kink in the $\phi^4$ model
\begin{equation}
L[\phi]=\int_{-\infty}^{\infty} \left( \frac{1}{2} (\partial_t\phi)^2 -
  \frac{1}{2} (\partial_x\phi)^2 - \frac{1}{2}(1-\phi^2)^2 \right) dx .
\label{Lagfield}
\end{equation}
The kink is a static solution of a Bogomolny equation with one continuous parameter $a$, or modulus, which fixes the position of the kink
\be
\Phi(x;a) = \tanh (x-a).
\ee
In general, a set of configurations $\Phi(x;X^i)$ is parametrized by a collection of moduli $X^i, i=1, 2,\dots N$. In the collective coordinate model (CCM) approach we insert it into the original Lagrangian and promote the moduli to time-dependent coordinates $X^i(t)$. Then, we integrate over the spatial dimensions and find the resulting CCM 
\be
L=T-V=\frac{1}{2} g_{ij}\dot{X}^i \dot{X}^j - V(X).
\ee
Here, $g_{ij}$ is the metric on the moduli space. It arises from the kinetic (temporal) part of the Lagrangian and reads
\begin{equation}
g_{ij}(X)=\int_{-\infty}^{\infty}  \frac{\partial {\Phi}
  (x;X)}{\partial X^i} \frac{\partial {\Phi} (x;X)}{\partial
  X^j}  dx \,.
\label{EffMet}
\end{equation}
$V$ is a potential on the moduli space
\begin{equation}
V(X)=\int_{-\infty}^{\infty} \left( \frac{1}{2} \Phi'(x;X)^2 
+ \frac{1}{2}(1-\Phi(x;X)^2)^2 \right) dx \,.
\end{equation}

In the case of the single kink, which is a trivial BPS solution, we obtain the simplest moduli space. It is a real line $\mathcal{M}=\mathbb{R}$ with a constant metric $g_{aa} =4/3$, which is exactly the mass of the kink $M$. The potential also has a constant value $V=M$. The resulting CCM reads 
\begin{equation}\label{lag1}
    L[a]=\frac{1}{2} M \dot{a}^2 - M,
\end{equation}
and describes a constant velocity non-relativistic motion of the kink. 

The $\phi^4$ kink possesses two bound modes that arise in linear perturbation theory. One zero mode $\eta_0(x;a)=1/\cosh^2(x-a)$, related to the translation invariance of the theory, and one vibrational mode, called the shape mode $\eta_1(x;a)=\sinh(x-a)/\cosh^2(x-a)$. The CCM model \eqref{lag1} is the simplest, as it includes only the trivial modulus, that is, only the zero mode.

Now we add the shape mode. This means that we have two moduli $(a,C)$ where $C$ is the amplitude of the shape mode. The corresponding set of configurations is
\begin{equation}
\Phi(x;a,C) = \tanh(x-a) + C \frac{\sinh(x-a)}{\cosh^2(x-a)} \,.
\label{kink-2dim}
\end{equation}
It gives rise to a two dimensional vibrational moduli space with a diagonal metric
\bea
g_{aa}&=&\frac{4}{3} +\frac{\pi}{2} C+\frac{14}{15}C^2 \,,  \\
g_{CC}&=&\frac{2}{3} . 
\eea
The off diagonal component vanishes due to the orthogonality of the linear modes. It is an important fact that the previously constant metric component $g_{aa}$ receives a non-trivial modification due to the vibrational mode. Such a coupling between the zero and massive mode is usually referred to as the Coriolis effect \cite{Raw}. 

It is not surprising that the effective potential in the resulting CCM also depends on $C$
\be
V(C)=\frac{4}{3} + C^2+\frac{\pi}{8} C^3 +\frac{2}{35}C^4.
\ee
This comes from the fact that the configurations (\ref{kink-2dim}) are only solutions of linearized field equations. Thus, their energy depends on $C$ \footnote{Strictly speaking, the amplitude of the mode is not the proper modulus. It does not parametrize the solutions leading to family of energetically equivalent states but rather parametrize field configurations which are believed to be important in some dynamical processes, e.g., scatterings. However, it is a commonly used terminology to call them modulus as well.}. The resulting CCM describes the vibrating kink in motion
\be
L[a,C]=\frac{1}{2}g_{aa}\dot{a}^2+\frac{1}{2}g_{CC}\dot{C}^2-V(C).
\label{vib-kink-CCM}
\ee

Naively, one could think that the modification of the moduli space metric $g_{aa}$ by the linear and quadratic term in $C$ is a small effect without any importance on the kink dynamics. However, this is a crucial modification for the correct description of the kink-antikink collisions within the corresponding CCM model \cite{MORW}. This originates from the observation that the vibrating kink CCM (\ref{vib-kink-CCM}) has a stationary solution $\dot{a}=v=\mbox{const.}$ and $C_0=C_0(v^2)=\mbox{const.}$, where $C_0$ follows from \cite{MORW}
\be
\frac{v^2}{2} \frac{d g_{aa}}{d C}  = \frac{dV}{dC}.
\ee
Thus, from the point of view of the CCM formulation, the kink with a constant velocity has a {\it non-zero} amount of the shape mode excited. This is of great importance for the proper specification of the initial conditions for the kink-antikink collision in the CCM \cite{MORW}. Otherwise, the CCM results do not correctly reproduce the famous chaotic pattern in the final state formation \cite{sug, CSW}. We also comment that this effect is intimately related with the perturbative restoration of the Lorentz covariance of the theory. Indeed, in the usual moduli space the kink moves in a completely non-relativistic manner. The metric modification due to the inclusion of the shape mode allows for an approximation of the Lorentz contraction and brings the CCM closer to the original Lorentz invariant model \cite{MORW}. 

In the subsequent analysis, we want to verify whether there is a similar modification of the moduli space of the 1-vortex.

\vspace*{0.2cm}

Because of the fact that in the single kink sector the moduli space metric and the effective potential do not depend on the trivial modulus $a$, they can be computed from the infinitesimal action of the zero mode. Indeed, let us consider an infinitesimal translation of the kink from $a=0$ to $a=\epsilon$
\be
\tanh(x-\epsilon)= \tanh(x) - \frac{\epsilon}{\cosh^2(x)} + o(\epsilon).
\ee
Now, the amplitude of the zero mode is taken as a modulus and promoted to a time-dependent coordinate. This gives exactly the same constant metric $g_{\epsilon \epsilon} = M$. The effective potential at $O(\epsilon^2)$ order is again $V(\epsilon)= M$. There are higher order corrections but they vanished in $
\epsilon \to 0$ limit. Thus, we reproduce \eqref{lag1}. This is not a surprising result. It just confirms the fact that the zero mode generates the motion along the $a$ direction.

Analogously, to obtain the vibrating kink CCM (\ref{vib-kink-CCM}), it is enough to consider the infinitesimal version of (\ref{kink-2dim}) 
\bea
\Phi(x;\epsilon, C) &=& \tanh(x) - \frac{\epsilon}{\cosh^2(x)} + C  \frac{\sinh(x)}{\cosh^2(x)} \nonumber
\\
&+&  \frac{\epsilon C }{\cosh(x)}  \left(1 - \frac{2}{\cosh^2(x)}  \right)+o(\epsilon). \label{kink-eA}
\eea
The resulting metric functions are
\bea
g_{\epsilon \epsilon}&=&\frac{4}{3}+\frac{\pi}{2}C+\frac{14}{15} C^2, \;\;\;
g_{CC}=\frac{2}{3} + \frac{14}{15} \epsilon^2, \nonumber \\
g_{\epsilon C}&=&\frac{\epsilon}{4}\left( \pi + \frac{56}{15}C \right),
\eea
and agree with the previous result in the limit $\epsilon\to 0$. The same happens for the effective potential. 

We remark that it is crucial to include the last term in (\ref{kink-eA}) which takes into account the mutual interaction of the shape and zero mode. The linear and quadratic mode amplitude modifications of the metric component $g_{\epsilon \epsilon}$ come from this term.

%%%%%%%%%%%%%%%%%%%%%%%%%%%%%%%%%%%%%%%%%%%%%%%%%%
\section{Unit charge vortex and its shape mode}
%%%%%%%%%%%%%%%%%%%%%%%%%%%%%%%%%%%%%%%%%%%%%%%%%%
Now we turn to the main topic of our work and consider the Abelian-Higgs  model in (2+1) dimensions
\begin{equation}\label{Eq:LagrangianDensity}
    \mathcal{L} = - \frac{1}{4}F_{\mu\nu}F^{\mu \nu} + \frac{1}{2}D_i\Phi \overline{D_i\Phi} + \frac{\lambda}{8}(1-\Phi\overline{\Phi})^2,
\end{equation}
where $A_\mu=(A_0,A_1,A_2)$ is the gauge potential and $\Phi$ is the Higgs complex scalar. Moreover, we define the covariant derivative $D_\mu = \partial_\mu - i A_\mu (x)$ and the electromagnetic field tensor $F_{\mu \nu}(x) = \partial_{\mu}A_\nu(x) -\partial_{\nu}A_{\mu}(x)$. The metric is $\eta^{\mu\nu} = \{+,-,-\}$. 

In the  temporal gauge $A_0=0$, the resulting field equations are 
\bea
\frac{1}{2} \partial_0^2 \Phi - \frac{1}{2} D_iD_i \Phi = -\frac{\lambda}{4} \Phi (\Phi \bar{\Phi}-1), \\
\frac{1}{2} \partial_0^2 A_j - \partial_k F_{kj} = - \frac{i}{2} \left[ \bar{\Phi} D_j \Phi - \Phi \overline{D_j \Phi} ,\right].
\eea
The fields must also obey the Gauss law constrain 
\be
\partial_i \dot{A}_i + \frac{i}{2} \left( \bar{\Phi} \partial_0 \Phi - \Phi \partial_0 \bar{\Phi}\right)=0. 
\ee

The static unit charge vortex $(\Phi^v, A^v_\mu)$ is an axially symmetric solution of the field equations. In the radial gauge $A_r=0$ and in  axial coordinates, it reads
\be
\Phi^v(r,\theta) = f(r) e^{i\theta}, \;\;\; A^v_\theta = \frac{\beta(r)}{r},
\ee
where the profile functions $f(r)$ and $\beta(r)$ obey 
\bea
\frac{d^2 f}{dr^2} +\frac{1}{r} \frac{df}{dr} -\frac{(1-\beta)^2}{r^2} f +\frac{\lambda}{2} f(1-f^2)&=&0,\label{ode21} \\
\frac{d^2 \beta}{dr^2} -\frac{1}{r} \frac{d \beta}{dr} +(1-\beta) f^2 &=&0.\label{ode22}
\eea

For the BPS limit, it can be proved that \eqref{ode21}-\eqref{ode22} can be rewritten in the following form using the Bogomolny decomposition
\bea
\frac{df}{d r}&=&\frac{f}{r}  [1-\beta],\label{ode11}\\
\frac{d\beta}{d r}&=&\frac{r}{2}[1-f^2] \, . \label{ode12}
\eea

The boundary conditions at  $r \to \infty$ are $f(r) \to 1 $ and $\beta (r) \to 1$ and guarantee the finite energy and nonzero topological charge. At $r\to 0$, both functions approach 0, that is, $f\sim d_0r$ and $\beta\sim c_0 r^2$ with $d_0,c_0$ being real numbers.

\begin{figure}
 \includegraphics[width=0.4\textwidth]{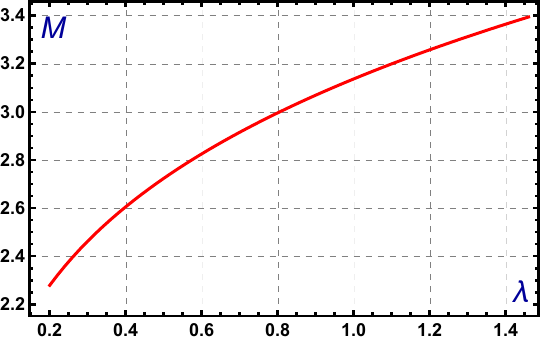}
     \includegraphics[width=0.4\textwidth]{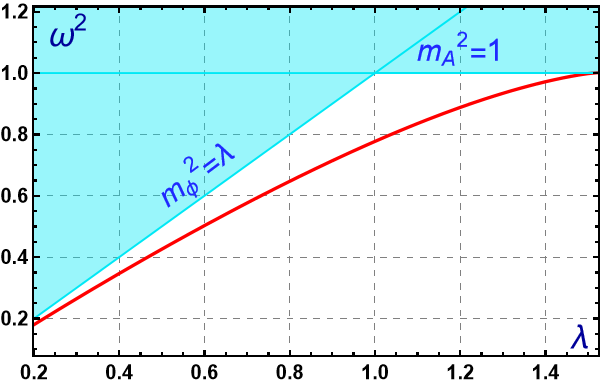}
    \caption{The mass (upper panel) and the spectrum (lower panel) of the 1-vortex as a function of the parameter $\lambda$. The discrete eigenvalue corresponds to the red curve, while the shaded blue area represents the continuous spectrum.
}
     \label{fig:shape_mode}
\end{figure}

\begin{figure}
     \includegraphics[width=0.4\textwidth]{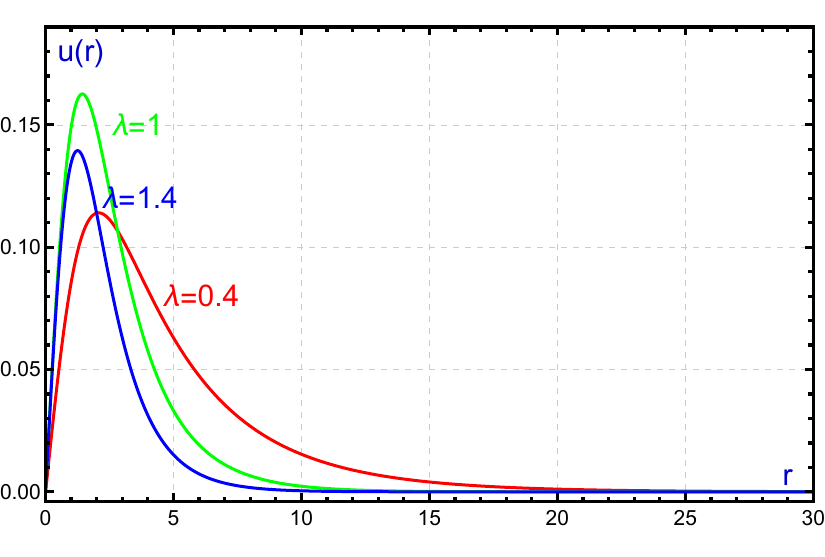}
      \includegraphics[width=0.4\textwidth]{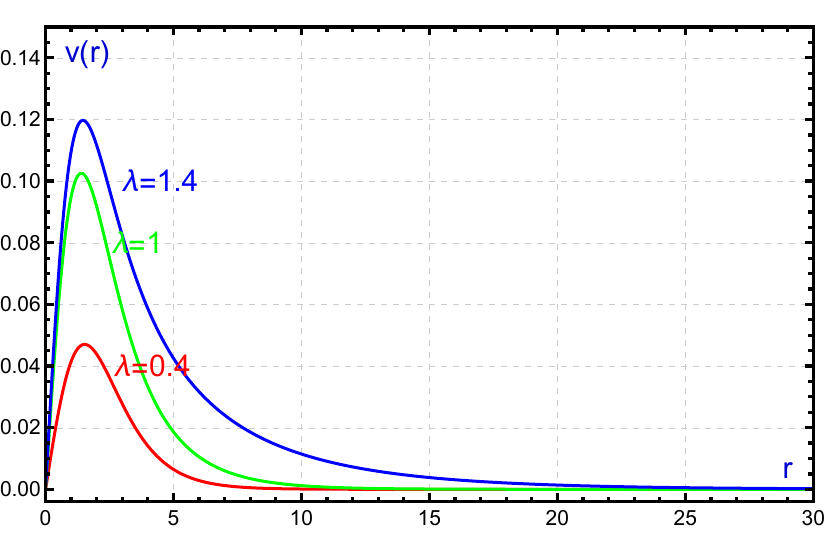}
    \caption{The normalized shape mode (the Higgs $u$ and the gauge field $v$ component) of the single vortex for $\lambda=0.4, 1, 1.4.$}
         \label{fig:profiles}
\end{figure}

It is a matter of fact that the 1-vortex solution $\Psi\equiv (\Phi^v_1,\Phi^v_2, A^v_1, A^v_2)$, hosts a shape mode. Recently, it has been shown that for $\lambda\lesssim1.5$ there is always a vibrating radial mode \cite{ID}. To find such a mode, a small perturbation
$\xi = (\eta_1,\eta_2,a_1,a_2 )^T$
is added to the unit vortex. The standard linear perturbation theory finds one radial mode with the following eigenvector 
\be
\xi(\vec{x}) = \left( 
\begin{array}{c}
     u(r)\cos \theta  \\
     u(r)\sin \theta \\
     v(r)\sin \theta \\
     -v(r)\cos \theta
\end{array}
\right), 
\ee
where functions $u(r), v(r)$ are solutions of two coupled second order differential equations forming a eigen-problem to the eigen-frequency $\omega$, which reads
\bea
 \omega^2 u &=& - \frac{d^2 u}{dr^2}- \frac{1}{r} \frac{d u}{dr}   \\ 
&+& \left( \frac{(1-\beta)^2}{r^2} +\frac{3\lambda}{2} f^2 - \frac{\lambda}{2} \right) u +\frac{2}{r} (1-\beta)fv, \nonumber \\
 \omega^2 v &=& - \frac{d^2 v}{dr^2}- \frac{1}{r} \frac{d v}{dr} \\
&+& \left( \frac{1}{r^2} +f^2 \right) v +\frac{2}{r} (1-\beta)f u, \nonumber
\eea
see \cite{Goodband1995,Kojo2007,AlonsoIzquierdo2024,ID} for details.  For $\lambda\approx1.5$ the shape mode hits the mass threshold and transmutes into a quasinormal mode, see Fig. \ref{fig:shape_mode}. In Fig. \ref{fig:profiles}, we show the shape mode for some values of $\lambda$. We always assume that the mode is normalized to 1. 

Higher charge vortices have a much more involved mode structure. Especially in the BPS case, the frequencies and even the number of modes can change as we pass through different BPS solutions, that is, as we flow on the moduli space. 

%%%%%%%%%%%%%%%%%%%%%%%%%%%%%%%%%%%%%%%%%%%%%%%%%%
\section{Moduli space of single vortex with shape mode}
%%%%%%%%%%%%%%%%%%%%%%%%%%%%%%%%%%%%%%%%%%%%%%%%%%
The construction of moduli space based on the BPS (multi)-vortex solutions is well known. In the first step, we divide the Lagrangian into two parts: kinetic $T$ and potential $V$ (we assume the temporal gauge)
\bea
T&=& \frac{1}{2}\int d^2x\left( \dot{\Phi} \dot{\bar{\Phi}} + \dot{A}_i \dot{A}_i\right), \label{g} \\
V&=& \frac{1}{2} \int d^2x \left( D_i \Phi \overline{D_i\Phi} + F_{12}^2 +\frac{\lambda}{4} (1-\Phi \bar{\Phi})^2  \right),
\eea
where $V$ is nothing but the static energy. The kinetic energy will give the metric on the moduli space provided $(\dot{\Phi}, \dot{A}_i)$ obeys the Gauss law. This is an important requirement, since the Gauss law is the condition which guarantees that $(\dot{\Phi}, \dot{A}_i)$ is orthogonal to the gauge orbit through $(\Phi, A)$. In its seminal paper Samols found the metric in the critical case, restricting $T$ to $(\dot{\Phi}, \dot{A}_i)$ satisfying the Gauss law and a linearized version of the Bogomolny equations \cite{Samols}. Finally, the potential $V$ is gauge invariant by construction and, of course, in the BPS limit it takes a constant value proportional to the topological charge.  

To compute the moduli space metric for a set of configurations where the shape mode is included,  we cannot directly use Samols' approach. In contrast, we will numerically compute the metric using the definition (\ref{g}). Again, in order to maintain the gauge invariance, the Gauss law will be imposed.  

An important observation is that, in gauge theories, the action of the spatial translation is not identical to the action of the corresponding zero mode. Indeed, for the static $N=1$ vortex, the naive infinitesimal translation along $x$ axis
\bea
\Phi^v(x_1-\epsilon, x_2) &=& \Phi^v (\vec{x}) - \epsilon \Phi^v_1(\vec{x}) + o(\epsilon), \\
A_1^v(x_1-\epsilon, x_2) &=& A_1^v(\vec{x}) -\epsilon \partial_1 A^v_1 (\vec{x})+ o(\epsilon),   \\
A_2^v(x_1-\epsilon, x_2) &=& A_2^v(\vec{x}) -\epsilon \partial_1 A^v_2 (\vec{x})+ o(\epsilon) , 
\eea
has to be replaced by 
\bea
T_{\epsilon} \Phi^v(x_1, x_2) &=& \Phi^v (\vec{x}) - \epsilon \left( \partial_1 \Phi^v(\vec{x}) - iA_1(\vec{x})\right)+ o(\epsilon) \nonumber \\
&=& \Phi^v (\vec{x}) - \epsilon D_1 \Phi^v(\vec{x}) + o(\epsilon),  \\
T_{\epsilon} A_1^v(x_1, x_2) &=& A^v_1(\vec{x}) + o(\epsilon),   \\
T_{\epsilon} A_2^v(x_1, x_2) &=& A^v_2(\vec{x}) -\epsilon \left( \partial_1 A^v_2 (\vec{x})- \partial_2 A^v_1 (\vec{x}) \right) + o(\epsilon) \nonumber \\
&=& A^v_2(\vec{x}) -\epsilon B^v(\vec{x}) + o(\epsilon),  
\eea
where $B^v(\vec{x})=F_{12}(\vec{x})$ and  where the terms proportional to $\epsilon$ can be put together in the  vector
\be
\xi_0=\left( D_1\Phi^v_1, D_1\Phi^v_2, 0, B^v \right)^T,
\ee
forming the corresponding zero mode previously found in \cite{Weinberg1979,Tong2014}. 

As usual in the collective coordinate framework, the time dependence is hidden in the evolution of the parameters (coordinates). Thus, 
\bea
\dot{\Phi} (x_1, x_2) &=& - \dot{\epsilon} D_1 \Phi^v(\vec{x}) + o(\epsilon), \\
\dot{A}_1(x_1, x_2) &=& o(\epsilon), \\
\dot{A}_2(x_1, x_2) &=& -\dot{\epsilon} B^v(\vec{x}) + o(\epsilon).  
\eea
These fields obey the Gauss law and, therefore, keep the perturbation in the perpendicular direction to the gauge orbit. 

Inserting  the translated field and their time derivatives into the Lagrangian \eqref{Eq:LagrangianDensity} we arrive at the simplest CCM, $L[\epsilon]=\frac{1}{2}g_{\epsilon \epsilon} \dot{\epsilon}^2 - M$, where the moduli space metric reads
\bea
g_{\epsilon \epsilon} &=& \int d^2x \left( |D_1\Phi^v|^2 +  (B^{v})^2 \right)  \nonumber \\
&=& \int d^2x \left( \frac{1}{2} |D_1\Phi^v|^2+\frac{1}{2} |D_2\Phi^v|^{2} + (B^{v})^2 \right)  \\
&=& V + \frac{1}{2} \int d^2 x \left( (B^{v})^2 - \frac{\lambda}{4} (1-|\Phi^v|^2)^2 \right) \rightarrow M.\nonumber \;\;
\eea
To obtain this result, we first used that the following integral identity is satisfied 
\be
I= \int d^2 x \left( (B^{v})^2 - \frac{\lambda}{4} (1-|\Phi^v|^2)^2 \right) =0.
\ee
In the critical case, the integrand follows from the second Bogomolny equation \eqref{ode12}. Outside the BPS limit, the integral vanishes due to the vanishing of the field momentum of the field. 
Next, we impose the limit $\epsilon \to 0$. This must be performed, as nothing depends on the value of this parameter. In summary, we find that the effective potential reduces to $M$. As expected, the resulting CCM describes a constant motion of the vortex.

In the next step, we include the shape mode $\xi$. Then, taking once again the gauge corrected infinitesimal translation along $x$-axis, we arrive at the following set of configurations which define the vibrational moduli space
of the single vortex up to $o(\epsilon)$ terms
\bea
\tilde{\Phi} &=& \Phi^v (\vec{x}) - \epsilon D_1 \Phi^v(\vec{x}) + C \eta (\vec{x})  - \epsilon C D_1 \eta(\vec{x}), \\
\tilde{A_1} &=& A_1^v(\vec{x}) + C a_1 (\vec{x}),  \\
\tilde{A_2} &=& A_2^v(\vec{x}) -\epsilon B^v(\vec{x}) +C a_2 (\vec{x})- \epsilon C B^s (\vec{x}).
\eea
Here, $C$ is the amplitude of the shape mode $\xi \equiv (\eta, a_1,a_2)^T$ and $B^s= \partial_1 a_2-\partial_2a_1$ is the magnetic field coming from the shape mode. 

This leads to the time derivative which should be inserted while we compute the corresponding CCM
\bea
\dot{\tilde{\Phi}} &=& - \dot{\epsilon} D_1 \Phi^v(\vec{x}) + \dot{C} \eta (\vec{x}) - \dot{\epsilon}C  D_1 \eta (\vec{x}),  \\
\dot{\tilde{A_1}} &=&  \dot{C} a_1 (\vec{x}),\\
\dot{\tilde{A_2}} &=& -\dot{\epsilon} B^v(\vec{x}) + \dot{C} a_2 (\vec{x})- \dot{\epsilon} C B^s (\vec{x}).
\eea
Here, we omit $\epsilon \dot{C}$ terms as they will vanish for $\epsilon \to 0$. It is easy to check that this configuration fulfills the Gauss law. 
Now we plug these expressions into the Lagrangian \eqref{Eq:LagrangianDensity}. We integrate over the spatial coordinates and take the limit $\epsilon \to 0$.  In the end, we replace the infinitesimal $\epsilon$ by the modulus $a$, which is just a finite (not necessarily infinitesimal) position of the vortex along the $x$-axis. The resulting CCM is
\be
L[a,C]=\frac{1}{2}g_{aa}\dot{a}^2 + \frac{1}{2}\dot{C}^2 - V(C).
\ee

\begin{figure}
     \includegraphics[width=0.4\textwidth]{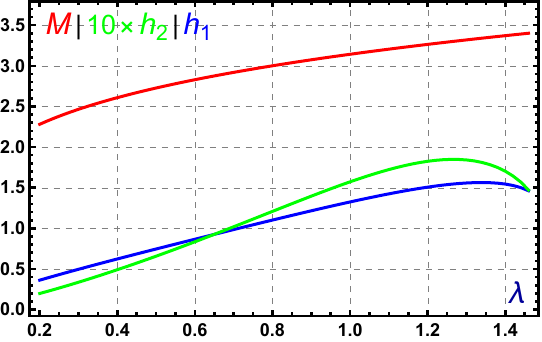}
    \includegraphics[width=0.4\textwidth]{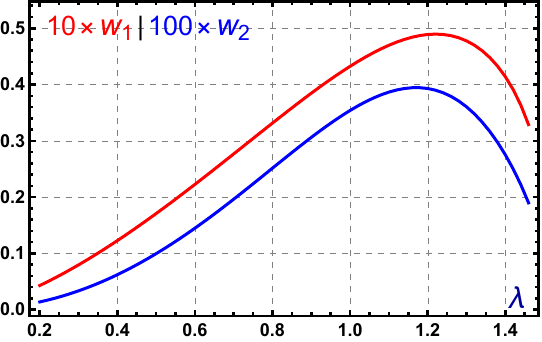}
    \caption{Coefficients in the 1-vortex CCM based on the shape mode.}
     \label{fig:ccm}
\end{figure}

The corresponding two-dimensional moduli space spanned by the coordinates $C$ and $a$ has the following diagonal metric 
\bea 
g_{aa} &=& M + h_1 C + h_2 C2, \\
g_{CC} &=& 1.
\eea
The constants $h_1, h_2$ give the modification of the flat single-vortex moduli space due to the normal mode. We compute them as functions of the coupling constant $\lambda$. The result is presented in Fig. \ref{fig:ccm} upper panel, where we also show $M(\lambda)$. We clearly see that $M \gg h_1 \gg h_2$. We also see that the functions $h_{1,2}$ take the maximum value close to $\lambda \approx 1.3$. Quite surprisingly, the self-dual limit is not distinguished \cite{AlonsoIzquierdo2024}.

The effective potential reads
\be
V(C)=M+\frac{1}{2}\omega^2 C^2 + w_1C^3+w_2C^4,
\ee
where the numerical coupling constants $w_1, w_2$ and the frequency of the shape mode $\omega$ are again functions of $\lambda$.  $w_1$ and $w_2$ are plotted in Fig. \ref{fig:ccm}, lower panel. Again, $\omega^2 \gg w_1 \gg w_2$. This means that terms of higher order in the amplitude of the shape mode are not only subleading due to the assumption that the amplitude $C$ is small but they are also multiplied by smaller coupling constants. 

\vspace*{0.2cm}

As in the case of the kink, this two-dimensional CCM has a stationary solution which describes a constant velocity motion of the vortex, $v=\dot{a}$. Due to the dependence of the metric on the amplitude of the mode, a nonzero velocity leads to a nonzero amplitude of the mode $C_0$, which is a solution of the following algebraic equation
\be
\frac{v^2}{2}\left(h_1+2h_2C_0\right)=\omega^2 C_0+3w_1C^2_0+4w_2C^3_0.
\ee
Using that $w_1$ and $w_2$ are small numbers and assuming that the amplitude of the mode is not too large,  we can omit these terms and find an approximate formula for the excitation of the shape mode
\be
C_0 \approx \frac{1}{2} \frac{h_1v^2}{\omega^2-h_2v^2} \approx  \frac{1}{2} \frac{h_1v^2}{\omega^2}.
\ee
We conclude that the moduli space metric of the single vortex receives a modification due to the excitation of the mode. Of course, this also holds for the vortices with higher topological charge. This means that the CCM proposed in \cite{AMMW} should also be improved by a similar modification of the moduli space metric. Then, the upper stationary solution defines the initial condition for the scattering of unexcited vortices. Generalization to excited vortices is straightforward. 
\begin{figure}
     \includegraphics[width=0.23\textwidth]{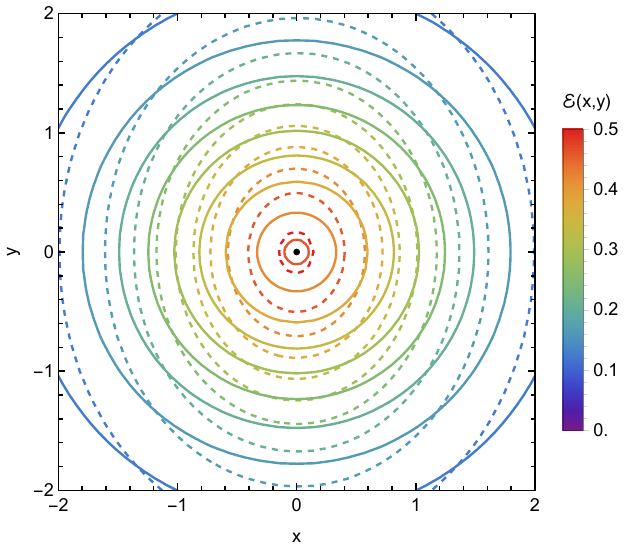}
      \includegraphics[width=0.23\textwidth]{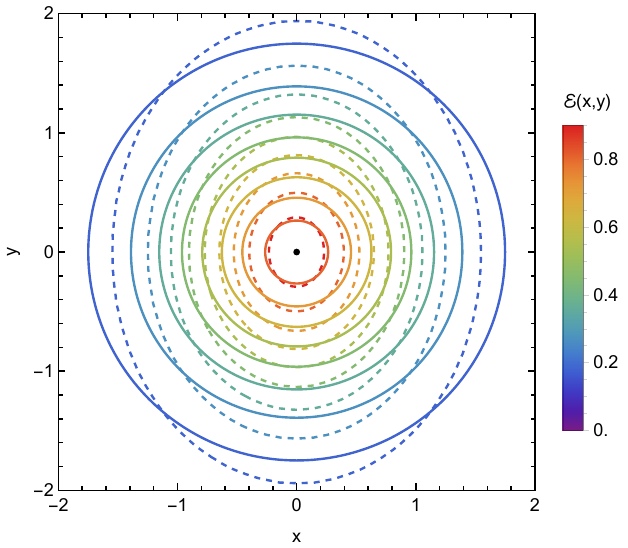}
            \includegraphics[width=0.23\textwidth]{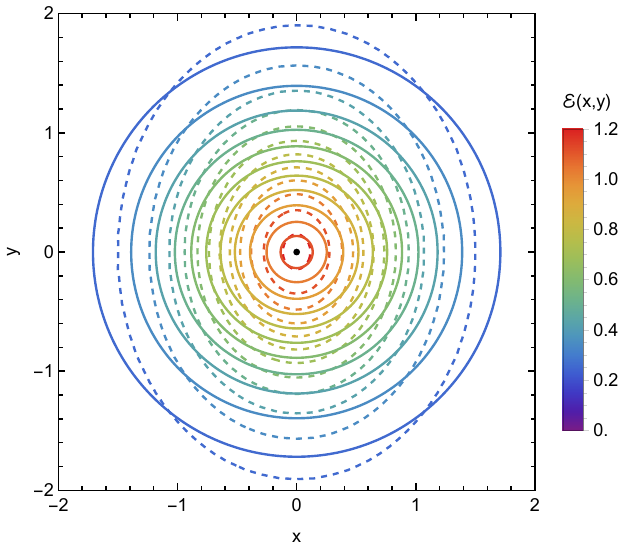}
    \caption{Comparison between  the energy density obtained from the collective coordinate method with the internal mode (solid lines) and the numerical simulation (dashed lines) for a vortex with $n=1$ and velocity $v=0.6$. Here $\lambda=0.4,1,1.4$.}
         \label{fig:energy_den_mode}
\end{figure}

For a one-dimensional kink, such a stationary solution approximately restores the Lorentz contraction of the moving soliton. However, here the situation is more subtle. The addition of the axially symmetric shape mode modifies the profile of the vortex in an axially symmetric way. This is not what happens with a boosted vortex. The axially symmetric vortex becomes squeezed in the direction of the motion, which leads to an elliptic shape, see Fig. \ref{fig:energy_den_mode}. Therefore, the inclusion of the radial shape mode is rather unsatisfactory to recover the (partial) Lorentz invariance at the level of the collective model.

%%%%%%%%%%%%%%%%%%%%%%%%%%%%%%%%%%%%%%%%%%%%%%%%%%
\section{Derrick Mode}
%%%%%%%%%%%%%%%%%%%%%%%%%%%%%%%%%%%%%%%%%%%%%%%%%%

It is well known that the Derrick mode \cite{R}, as well as its higher-order generalizations \cite{AMORW}, play a distinguished role among various deformations of a soliton.  Derrick modes arise due to the scale transformation $x^i \to \lambda^i x^i$ (no summations over the indices) and reflect the ability of the soliton to be compressed or expanded. Because of that, they naturally serve as degrees of freedom, which may effectively approximate the Lorentz contraction of a boosted soliton. Interestingly, in the case of kinks, the Derrick mode often reproduces very well the first vibrational mode (shape mode) \cite{AMORW}. This was observed, e.g., in the $\phi^4$, the Christ-Lee and the double sine-Gordon model \cite{ACORW}. Thus, to model the multi-kink scatterings, the CCMs based on the Derrick mode or the shape mode serve equally well. In fact, the inclusion of the higher Derrick modes through the Perturbative Relativistic Moduli Space framework, leads to a CCM which predicts some observables with an exceptionally good precision \cite{ACORW} (e.g., the critical velocity which separates the single backward scattering from multi-bounce windows or annihilation).  

Here, we begin with the effect of the scaling transformation {\it only} in the $x_1$ direction, $x_1 \to (1+C)x_1$, which applied to the unite charge vortex it gives 
\begin{equation}
    \begin{split}
        T_\delta \Phi^v(x_1,x_2) &= \Phi^v (\vec{x}) + C x_1 D_1\Phi^v (\vec{x}), \\
         T_\delta  A^v_1(x_1,x_2) &= A^v_1(\vec{x}),\\
         T_\delta  A^v_2(x_1,x_2) &= A^v_2(\vec{x}) + C x_1B^v(\vec{x}),
    \end{split}
\end{equation}
where terms up to linear order in $C$ are kept. As in the previous construction, the naive result must be modified to obey the Gauss law. Combining these deformations with the zero mode translation along the $x$-axis we arrive at the following set of configurations
\begin{equation}
    \begin{split}
        \tilde{\Phi} &= \Phi^v(\vec{x}) + \epsilon \,D_1 \Phi^v(\vec{x}) + C x_1 D_1 \Phi^v(\vec{x}) \\
        &+ \epsilon C D_1(x_1 D_1\Phi^v(\vec{x})) ,\\
        \tilde{A}_1 &= A^v_1(\vec{x}),\\
        \tilde{A}_2&= A^v_2(\vec{x}) + \epsilon \, B^v(\vec{x}) + C x_1 B^v(\vec{x}) + \epsilon C \partial_1(x B^v(\vec{x})) .
    \end{split}
\end{equation}
As required, the fields obey the Gauss law and keep the perturbation in the perpendicular direction to the gauge orbit.

\begin{figure}
     \includegraphics[width=0.4\textwidth]{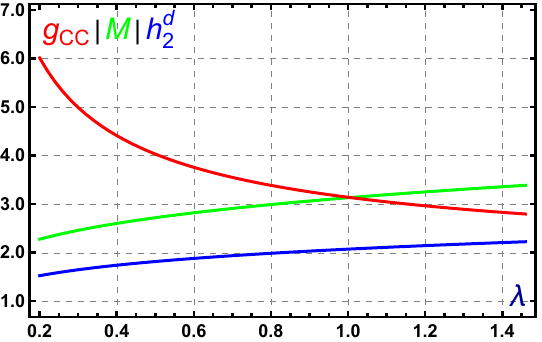}
    \includegraphics[width=0.4\textwidth]{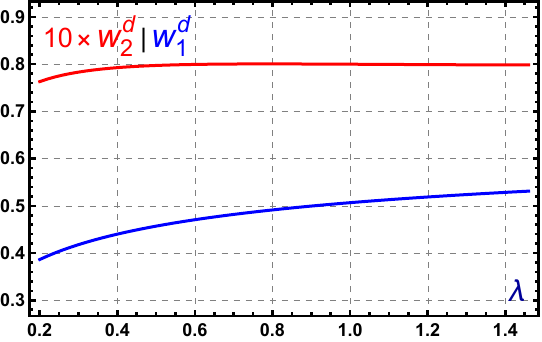}
    \caption{Coefficients in the $x$-Derrick mode based effective model for the 1-vortex.}
     \label{fig:ccm_der}
\end{figure}

This set of configurations gives rise to the following $x$-Derrick mode based CCM
\be
L_d[a,C] = \frac{1}{2} g_{aa} \dot{a}^2 + \frac{1}{2}g_{CC}\dot{C}^2 - V(C).
\ee
The metric functions have the same structure as before
\be
g_{aa} = M + h_1^d C + h_2^d C^2. 
\ee
One can show that $h_1^d = M$. In addition, $g_{CC}$ is not 1 which reflects the fact that the assumed form of the Derrick mode is not normalized to 1. In Fig. \ref{fig:ccm_der} upper panel we plot $h_2^d$ and $g_{CC}$.  
\\
The potential is again a fourth order function
\be
V(C)=M+\frac{1}{2} w_0^d C^2 + w_1^dC^3+w_2^dC^4.
\ee
Again, one can show that $w_0^d=M$. In Fig. \ref{fig:ccm_der} lower panel we plot the couplings $w_1^d$ and $w_2^d$ as functions of $\lambda$. 

This CCM also possesses a stationary solution $\dot{a}=v$ and $C=C_0$ where $C_0$  obeys the following equation
\be
\frac{v^2}{2}\left(M+2h_2^dC_0\right)=M C_0+3w_1^dC^2_0+4w_2^dC^3_0,
\ee
which, at quadratic order in velocity, is just $C_0=v^2/2$. Thus, a constant velocity motion of the 1-vortex requires a nonzero excitation of the Derrick mode. 

\begin{figure}
     \includegraphics[width=0.23\textwidth]{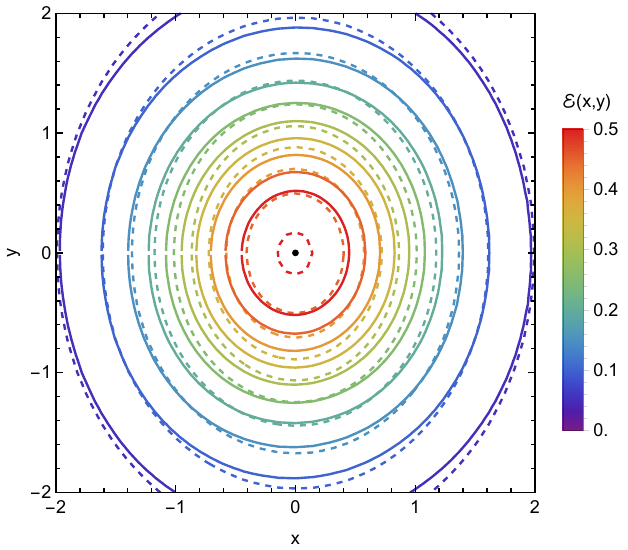}
      \includegraphics[width=0.23\textwidth]{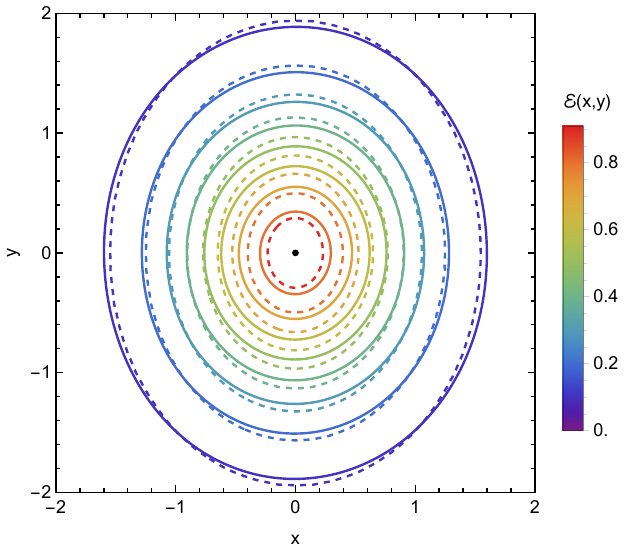}
            \includegraphics[width=0.23\textwidth]{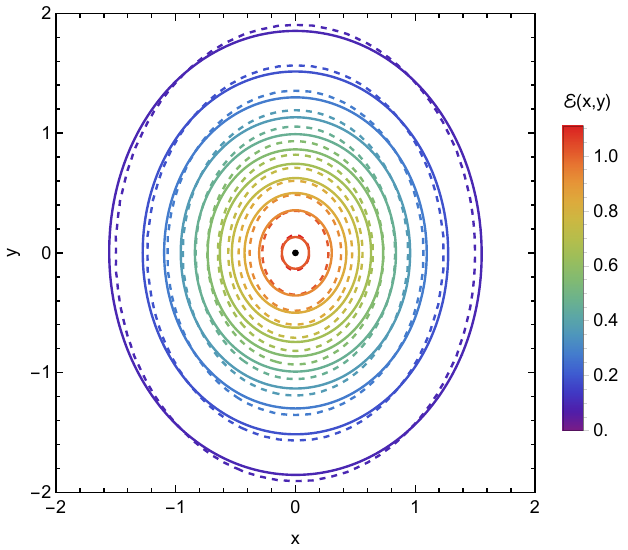}
    \caption{Comparison between  the energy density obtained from the $x$-Derrick mode CCM with (solid lines) and the numerical simulation (dashed lines) for a vortex with $n=1$ and velocity $v=0.6$. Here $\lambda=0.4,1,1.4.$}
         \label{fig:energy_den_mode_derr}
\end{figure}

\begin{figure}
    \centering
    \includegraphics[width=0.4\textwidth]{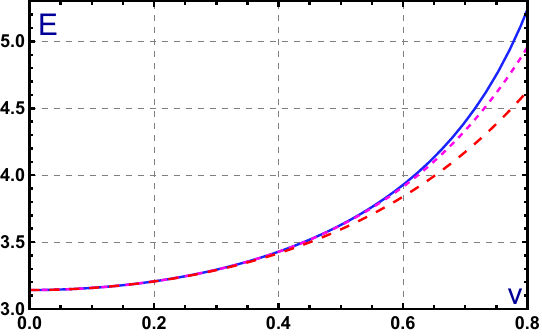}
    \caption{Comparison between the energy of the moving 1-vortex (solid line) with the energy of solution obtained in the CCM based on the $x$-Derrick mode (dotted line) and the shape mode (dashed line). Here $\lambda=1$.}
    \label{fig:energy}
\end{figure}

In the next step, we use the stationary solution of this CCM and plot the resulting field configuration. This is the static vortex with the addition of the $x$-Derrick mode. As shown in Fig. \ref{fig:energy_den_mode_derr}, it reproduces the Lorentz contraction of the moving vortex with very good precision. The agreement is significantly better than in the case of the shape-mode-based CCM. The same applies to the energy of the moving vortex. The CCM based on the $x$-Derrick mode also gives much better approximation to the relativistic energy of the single vortex than the shape-mode-based CCM, see Fig. \ref{fig:energy}, where we plot dependence of the energies on velocity for $\lambda=1$.  

\begin{figure}
    \centering
    \includegraphics[width=0.4\textwidth]{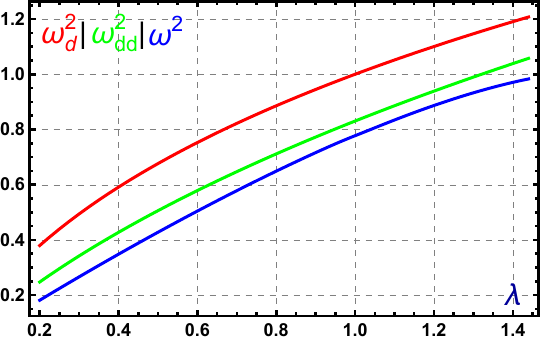}
     \caption{Comparison of the frequency of the shape mode $\omega$ and the frequencies of the radial-Derrick mode $\omega_{dd}$ and the $x-$Derrick mode $\omega_d$.}
    \label{fig:comp_freq}
\end{figure}

In the quadratic approximation the frequency of the $x$-Derrick mode reads
\be
\omega_d^2=\frac{M}{g_{CC}}.
\ee
In Fig. \ref{fig:comp_freq} we compare the frequency of the $x$-Derrick mode $\omega_d$ and the frequency of the shape mode $\omega$. Not surprisingly, the agreement is not perfect. The shape mode is a radially symmetric excitation, while the $x$-Derrick mode violates this symmetry. We found an approximately constant difference $\omega^2_d-\omega^2 \approx 0.2$. Note that the Derrick mode exists for any $\lambda$. This is not true for the bound mode, which does not exist for $\lambda>1.5$. However, for bigger $\lambda$, the shape mode does not completely disappear. It transmutes into a Feschbach resonance \cite{Feshbach1958,Mittleman1966}, which, at least for not too large $\lambda$, is a rather stable excitation. Hence, it still makes sense to approximate it (its real oscillating part) in terms of a Derrick mode. This type of resonance has revealed a notable impact on soliton dynamics in one and more spatial dimensions \cite{Forgacs2004,Russell2011,MartinCaro2024}.

\vspace*{0.2cm}

It is clear how this CCM should be further developed in order to correctly approximate the Lorentz contraction and, at the same time, be able to model the axially symmetric shape mode. The second Derrick deformation, the one along the $y$-axis, has to be added. This amounts to the following set of the configurations
\begin{equation}
    \begin{split}
        \tilde{\Phi} &= \Phi^v(\vec{x}) + \epsilon \,D_1 \Phi^v(\vec{x}) \\
        &+ C_1 x_1 D_1 \Phi^v(\vec{x})
+ \epsilon C_1D_1(x_1 D_1\Phi^v(\vec{x})) \\
&+ C_2 x_2 D_2 \Phi^v(\vec{x})
+ \epsilon C_2 D_1(x_2 D_2\Phi^v(\vec{x})) ,\\
        \tilde{A}_1 &= A^v_1(\vec{x}) - C_2 x_2 B^v(\vec{x}),\\
        \tilde{A}_2&= A^v_2(\vec{x}) + \epsilon \, B^v(\vec{x}) + C_1 x_1 B^v(\vec{x})\\
        &+ \epsilon\, C_1 \partial_1( x_1 B^v(\vec{x})) + \epsilon\, C_2 \partial_2(x_2 B^v(\vec{x})). 
    \end{split}
\end{equation}
Again, for simplicity reasons, we assumed that the vortex motion is along the $x$ axis. 

The new CCM based on the zero mode and both Derrick modes is written as
\begin{equation} \label{eq:CCM_2Derrick}
    \begin{split}
        L_{dd}[a, C_1,C_2] &= \dfrac{1}{2}g_{aa}\dot{a}^2+ \dfrac{1}{2}g_{C_1 C_1}\dot{C}_1^2 + \dfrac{1}{2}g_{C_2 C_2}\dot{C}_2^2\\
        &+ \dfrac{1}{2}g_{C_1 C_2}\dot{C}_1\dot{C}_2 - V(C_1,C_2).
    \end{split}
\end{equation}
The metric on the three-dimensional moduli space now has a more complex form
\be
    \begin{split} \label{eq:Metric2Derrick_AC1C2}
        g_{aa} &= M + h_{1,1}^{dd} C_1 + h_{1,2}^{dd} C_2\\
        &+ h_{2,1}^{dd} C_1^2 + h_{2,2}^{dd} C_2^2 + h_3^{dd} C_1C_2.     
    \end{split}
\ee
\begin{figure}
    \includegraphics[width=0.41\textwidth]{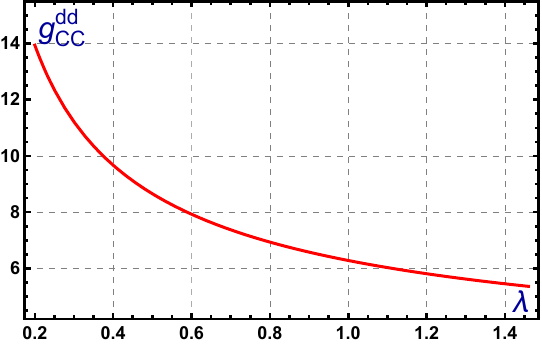}
    \includegraphics[width=0.4\textwidth]{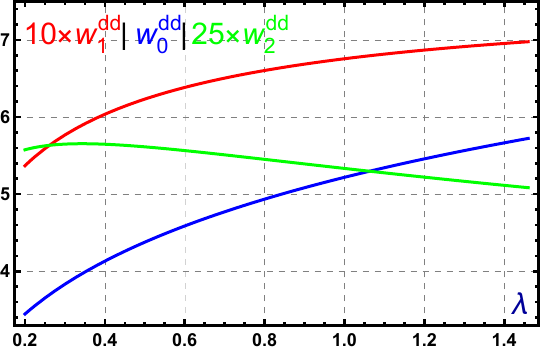}
    \caption{Coefficients in the radial-Derrick mode based effective model for the 1-vortex for a static configuration.}
     \label{fig:ccm_der2}
\end{figure}
The main difference from the previous effective models is that the metric exhibits non-diagonal terms reflecting the fact that the $x$ and $y$ scale deformations do not give rise to orthogonal modes. One can show that $g_{C_1C_1} = g_{C_2C_2}$. The potential $V(C_1,C_2)$ is once again a quartic polynomial in the mode amplitudes. 

For our purpose of reproducing the frequency of the shape mode, we do not display the coefficients in the metric and in the potential in full generality. 
In contrast, we assume an axially symmetric situation where $C_1 = C_2 = C$. This is the consistent assumption for the 1-vortex whose center of mass does not move, $\dot{a}=0$. The CCM model (\ref{eq:CCM_2Derrick}) is then reduced to 
\begin{equation} \label{eq:CCM_RadialDerrick}
    \begin{split}
        L_{dd}[C] &= \dfrac{1}{2}g_{CC}^{dd}\dot{C}^2 - V(C).
    \end{split}
\end{equation}
with
\begin{equation}
    V(C) = M + \dfrac{1}{2}w_{0}^{dd}C^2 + w_1^{dd} C^3 + w_2^{dd} C^4.
\end{equation}
In Fig. \ref{fig:ccm_der2}, we plot the metric coefficient (upper panel) and the potential coefficients (lower panel). In the quadratic approximation, this CCM provides the frequency of the radial Derrick mode as $\omega_{dd}=\sqrt{w_0^{dd}/g_{CC}^{dd}}$. This frequency is significantly closer to the frequency of the shape mode than in the case of $x$-Derrick mode, see Fig. \ref{fig:comp_freq}. Now, the difference between these frequencies is reduced to an approximately constant value, $\omega_{dd}^2 - \omega^2 \approx 0.06$. 

As expected, the inclusion of the Derrick modes in the $x$ and $y$ axes captures the axially symmetric shape mode and is able to approximate the Lorentz contraction. 

%%%%%%%%%%%%%%%%%%%%%%%%%%%%%%%%%%%%%%%%%%
\section{Summary}
%%%%%%%%%%%%%%%%%%%%%%%%%%%%%%%%%%%%%%%%%%
In the present work, we studied the impact of the massive vibrational modes on the moduli space metric in the case of the 1-vortex. Such modes do not only lead to the appearance of a mode-dependent effective potential (which consequently amounts to a mode-generated force) but also deform the metric component related to the zero mode, which is the change of the position of the 1-vortex. This has an important consequence if one uses the collective coordinate model (CCM) to reproduce the dynamics of vortices. Namely, on the level of the CCM, a constant motion of the 1-vortex requires a non-zero amount of the mode. 

This was tested in the CCMs based on a single mode: the shape mode and the $x$-Derrick mode. The shape mode is the proper bound mode of the 1-vortex and is a natural candidate for a CCM description of the vortex dynamics, see \cite{AMMW, AMMRW}. Unfortunately, both CCMs have some drawbacks. The shape-mode-based CCM does not reproduce the Lorentz contraction of the vortex, while the CCM based on the $x$-Derrick mode only crudely approximates the frequency of the shape mode. 

This problem is solved in the three-dimensional CCM (\ref{eq:CCM_2Derrick}) that contains two Derrick modes, that is, two scale deformations in the $x$ and $y$ directions. As a consequence, this CCM defines the appropriate initial conditions for the (excited) vortex-vortex and vortex-antivortex scattering in an effective theory, especially at relativistic velocities. This may also explain why, even in the BPS limit, in the highly relativistic scattering of initially unexcited 1-vortices we find vibrating solitons in the final state. A detailed investigation of this matter is left for future research. 

Our approach can also be applied to the collective description of the scattering of the BPS monopoles \cite{Max1, Max2}, especially if they are excited. We expect a similar modification of the metric functions, even though the normal vibrational modes should be replaced by quasinormal modes (Feshbach resonances) \cite{Forgacs2004, Russell2011}.

%%%%%%%%%%%%%%%%%%%%%%%%%%%%%%%%%%%%%%%%%%
\acknowledgments
%%%%%%%%%%%%%%%%%%%%%%%%%%%%%%%%%%%%%%%%%%

D.M.C., S.N.O. and A.W. acknowledge support from the Spanish Ministerio de Ciencia e Innovacion (MCIN) with funding from the European Union NextGenerationEU (Grant No. PRTRC17.I1) and the Consejeria de Educacion from JCyL through the QCAYLE project, as well as the grant
PID2023-148409NB-I00 MTM. D.M.C. and S.N.O. acknowledge financial support from the European Social Fund, the Operational Programme of Junta de Castilla y Le\'on and the regional Ministry of Education. Financial support of the Department of Education, Junta de Castilla y Le\'on, and
FEDER Funds is gratefully acknowledged (CLU-2023-1-05).


\newpage 

\begin{thebibliography}{99}


\bibitem{NO} H. Nielsen and P. Olesen, {\it Vortex-line models for dual strings}, Nuclear Physics B 61 (1973) 45.

\bibitem{Manton2002}
N.S. Manton and P. Sutcliffe, \textit{Topological solitons}, Cambridge University Press (2002).


\bibitem{Vilenkin2000}
A. Vilenkin and E.P.S. Shellard, \textit{Cosmic strings and other topological defects}, Cambridge University Press (2000).



\bibitem{BOG} E.~B. Bogomolny, {\it The stability of classical
solutions}, Sov. J. Nucl. Phys. {\bf 24} (1976) 449.

\bibitem{T} C.~H. Taubes, {\it Arbitrary $N$-vortex solutions to
the first order Ginzburg--Landau equations},
Commun. Math. Phys. {\bf 72} (1980) 277.

\bibitem{S}  D. Stuart, {\it Interaction of superconducting vortices and asymptotics of the Ginsburg-Landau flow}, Appl. Math. Lett. {\bf 9}(5) (1996) 27.

\bibitem{Samols} T.~M. Samols, {\it Vortex scattering}, Commun. Math. Phys. {\bf 145} (1992) 149.

\bibitem{NM} N. S. Manton, {\it A remark on the scattering of BPS monopoles}, Phys. Lett. {\bf B110} (1982) 54.

\bibitem{SR} E.~P.~S. Shellard and P.~J. Ruback, {\it Vortex scattering
in two-dimensions}, Phys. Lett. {\bf B209} (1988) 262. 

\bibitem{Ruback} P.~J. Ruback, {\it Vortex string motion in the
Abelian Higgs model}, Nucl. Phys. {\bf B296} (1988) 669.

\bibitem{Reb} C. Rebbi, R. Strilka and E. Myers, {\it Vortex scattering
in the Abelian Higgs model},
Nucl. Phys. B Proc. Suppl. {\bf 26} (1992) 240.

\bibitem{Sp} J. M. Speight, {\it Static intervortex forces}, Phys. Rev. {\bf D55} (1997) 3830.

\bibitem{AGM} A. Alonso Izquierdo, W. Garcia Fuertes
and J. Mateos Guilarte, {\it Dissecting zero modes and bound
states on BPS vortices in Ginzburg--Landau superconductors},
JHEP {\bf 05} (2016) 074.

\bibitem{AGMM} A. Alonso Izquierdo, W. Garcia Fuertes, N.S. Manton and
  J. Mateos Guilarte, {\it Spectral flow of vortex shape modes over
    the BPS 2-vortex moduli space},
    J. High Energy Phys. {\bf 01} (2024) 020.

\bibitem{AMMRW} A. Alonso Izquierdo, N. S. Manton,  J. Mateos Guilarte, M. Rees, A. Wereszczynski, {\it Dynamics of Excited BPS 3-Vortices}, arXiv:2502.15087.

\bibitem{AMMW} A. Alonso Izquierdo, N. S. Manton,  J. Mateos Guilarte, A. Wereszczynski, {\it Collective Coordinate Models for 2-Vortex Shape Mode Dynamics}, Phys. Rev. {\bf D110} (2024) 085006; arXiv:2405.20249.  
  
\bibitem{KRW} S. Krusch, M. Rees, T. Winyard, {\it Scattering of Vortices with Excited Normal Modes}, Phys. Rev. {\bf D110} (2024) 056050; arXiv:2406.04164.

\bibitem{AMRW} A. Alonso Izquierdo, J. Mateos Guilarte, M. Rees, A. Wereszczynski, {\it Spectral wall in collisions of excited Abelian Higgs vortices}, Phys. Rev. {\bf D110} (2024) 065004; [arXiv:2406.05725]. 

\bibitem{Raw} J. I. Rawlinson, {\it Coriolis terms in Skyrmion Quantization}, Nucl. Phys. {\bf B949} (2019) 114800. 

\bibitem{MORW} N. S. Manton, K. Oles, T. Romanczukiewicz, and A.
Wereszczynski, {\it Collective coordinate model of kink-antikink collisions in $\phi^4$ theory}, Phys. Rev. Lett. {\bf 127} (2021) 071601.

\bibitem{sug}  T. Sugiyama, {\it Kink-antikink collisions in the two-dimensional $\phi^4$ model}, Prog. Theor. Phys. {\bf 61} (1979) 1550.

\bibitem{CSW}  D. K. Campbell, J. F. Schonfeld, and C. A. Wingate,
{\it Resonance structure in kink-antikink interactions in $\phi^4$
theory}, Physica {\bf D9} (1983) 1.

\bibitem{ID} A. Alonso-Izquierdo, D. Miguelez-Caballero, {\it Dissecting normal modes of vibration on vortices in Ginzburg-Landau superconductors}, Phys. Rev. {\bf D110} (2024) 125026. 

\bibitem{Goodband1995}
M. Goodband and M. Hindmarsh, \textit{Bound states and instabilities of vortices}, Phys. Rev. \textbf{D52} (1995) 4621.

\bibitem{Kojo2007}
T. Kojo, H. Suganuma and K. Tsumura, \textit{Peristaltic modes of a single vortex in the Abelian Higgs model}, Phys. Rev. \textbf{D75} (2007) 105015.

\bibitem{AlonsoIzquierdo2024}
A. Alonso-Izquierdo, J.J. Blanco-Pillado, D. Migu\'elez-Caballero, S. Navarro-Obreg\'on and J. Queiruga, \textit{Excited Abelian-Higgs vortices: Decay rate and radiation emission}, Phys. Rev. \textbf{D110} (2024) 065009.


\bibitem{Weinberg1979}
E. J. Weinberg, \textit{Multivortex solutions of the Ginzburg-Landau equations},
Phys. Rev. \textbf{D19} (1979) 3008.  

\bibitem{Tong2014} 
D. Tong and K. Wong, \textit{Vortices and impurities}, J. High Energy Phys. \textbf{2014} (2014) 90. 

\bibitem{R} M. J. Rice, {\it Physical dynamics of solitons}, Phys. Rev. {\bf B28} (1983) 3587.

\bibitem{AMORW} C. Adam, N. Manton, K. Oles, T. Romanczukiewicz, A. Wereszczynski, {\it Relativistic Moduli Space for Kink Collisions}, Phys. Rev. {\bf D105} (2022) 065012.

\bibitem{ACORW} C. Adam, D. Ciurla, K. Oles, T. Romanczukiewicz, A. Wereszczynski, {\it Relativistic Moduli Space and critical velocity in kink collisions}, Phys. Rev. {\bf E108} (2023) 024221.

\bibitem{Feshbach1958}
H. Feshbach, \textit{Unified theory of nuclear reactions}, Annals of Phys. \textbf{5} (1958) 357.

\bibitem{Mittleman1966}
M. H. Mittleman, \textit{Resonances in multichannel scattering}, Phys. Rev. \textbf{147} (1996) 73.

\bibitem{Forgacs2004}
P. Forgacs and M. S. Volkov, \textit{Resonant excitations of the 't Hooft-Polyakov monopole}, Phys. Rev. Lett. \textbf{92} (2004) 151802.

\bibitem{Russell2011}
K. M. Russell and B. J. Schroers, \textit{On resonances and bound states of the 't Hooft-Polyakov monopole}, Phys. Rev. \textbf{D83} 
(2011) 065004.

\bibitem{MartinCaro2024} A. Garc\'ia Mart\'in-Caro, J. Queiruga and A. Wereszczynski, \textit{Feshbach resonances and dynamics of BPS solitons}, arXiv:2501.02589v2.

\bibitem{Max1} M. Bachmaier, G. Dvali, J. S. Valbuena-Bermudez, and M. Zantedeschi, {\it Confinement slingshot and gravitational waves}, Phys. Rev. {\bf D110} (2024) 016001.

\bibitem{Max2} M. Bachmaier, G. Dvali, J. Seitz, J. S. Valbuena-Bermudez, {\it Simulations of Magnetic Monopole Collisions}, arXiv:2502.01756.


\end{thebibliography}
\end{document}